\documentclass[aps,prb,amsfonts,amssymb,twocolumn,amsmath,preprintnumbers,nofootinbib,floatfix,showpacs]{revtex4}

\usepackage{bm}
\usepackage[dvips]{graphics}
\usepackage{graphicx}

\begin{document}

\title{Proximity-driven source of highly spin-polarized ac current on the basis of
superconductor/weak ferromagnet/superconductor voltage-biased
Josephson junction.}

\author{A. M. Bobkov}
\affiliation{Institute of Solid State Physics, Chernogolovka,
Moscow reg., 142432 Russia}
\author{I. V. Bobkova}
\affiliation{Institute of Solid State Physics, Chernogolovka,
Moscow reg., 142432 Russia}

\date{\today}

\begin{abstract}
We theoretically investigate an opportunity to implement a source
of highly spin-polarized ac current on the basis of
superconductor/weak ferromagnet/superconductor (SFS)
voltage-biased junction in the regime of essential proximity
effect and calculate the current flowing through the probe
electrode tunnel coupled to the ferromagnetic interlayer region.
It is shown that while the polarization of the dc current
component is generally small in case of weak exchange field of the
ferromagnet, there is an ac component of the current in the
system. This ac current is highly spin-polarized and entirely
originated from the non-equilibrium proximity effect in the
interlayer. The frequency of the current is controlled by the
voltage applied to SFS junction. We discuss a possibility to
obtain a source of coherent ac currents with a certain phase shift
between them by tunnel coupling two probe electrodes at different
locations of the interlayer region.
\end{abstract}

\maketitle

Various proximity and transport phenomena in hybrid structures
containing superconducting and ferromagnetic elements are
currently in the spotlight. The equilibrium transport and
proximity effect in such structures have been theoretically and
experimentally investigated recently in details as for the case of
weak ferromagnetic alloys (see Ref.\cite{buzdin} and references
therein) so as for half-metals like $CrO_2$
\cite{Eschrig03,Keizer06}.

Spin-dependent properties of nonequilibrium systems are also
actively investigated. In particular, the spin imbalance induced
in a ferromagnet (F) in nonequilibrium conditions was studied in
ferromagnet-superconductor-ferromagnet (FSF) junctions
\cite{takahashi99,maekawa02,tserkovnyak02,johansson04,pena05,morten05}.
It was found that antiferromagnetic alignment of the exchange
fields of the ferromagnets strongly suppresses superconductivity,
which leads to large magnetoresistive effect. The influence of the
interplay between Andreev reflection at the interface and spin
accumulation close to the interface was investigated
\cite{belzig00}. The effect of spin injection in Josephson
junctions was considered \cite{takahashi01}. The papers
\cite{morten05,morten04} have theoretically studied the spin
relaxation due to spin-flip scattering.

The authors of \cite{giazotto05,giazotto07} propose the
possibility of manipulating magnetization of a mesoscopic normal
(N) region through the Zeeman splitting of superconducting density
of states and applied voltage in voltage biased SNS or FSNSF
tunnel junctions. Also, the spin-polarized transport through
superconductor-normal metal hybrid structures, where the density
of states in a superconductor is Zeeman-splitted, was investigated
\cite{giazotto05,giazotto07,giazotto07a}. It was shown that, in
alternative to half-metallic ferromagnets, such junction can be
used to generate highly spin-polarized currents, which are tunable
in magnitude and sign by the bias voltage and exchange field. In
\cite{giazotto05,giazotto07} the normal metal region between the
superconducting leads has been considered to be long enough in
order to suppress the proximity effect and, consequently, ac
Josephson effect in it. However, if the junction length is of the
order of superconducting coherence length $\xi=\sqrt{\hbar
D/\Delta}$, where $D$ is the diffusion constant and $\Delta$ is
the superconducting order parameter in the leads, the interplay
between the proximity effect and the Zeeman exchange field in the
interlayer can lead to a number of novel qualitative effects. In
particular, it was shown that by tunnel coupling of the normal
regions of two superconductor-normal metal-ferromagnet trilayers
the absolute spin-valve effect can be realized for a certain
interval of voltages applied between the normal regions
\cite{hernando02}. For voltage-biased superconductor-weak
ferromagnet-superconductor Josephson junction the interplay
between the proximity effect and the Zeeman exchange field results
in additional peak-like features in the I-V characteristics
\cite{bobkovy06}. The ac Josephson current in
superconductor-ferromagnet-superconductor voltage biased junction
was also studied \cite{hikino08}.

In the present paper we study voltage-biased superconductor-weak
ferromagnet-superconductor (SFS) Josephson junction focusing on
the short enough interlayer (which is considered to be of the
order of superconducting coherence length). In this regime the
proximity effect between ferromagnet and superconductors is
essential. One of the well-known manifestations of the proximity
effect, which already takes place in equilibrium, is the so-called
minigap in the local density of states (LDOS) of the interlayer
\cite{mcmillan68,Fazio99}. Another important manifestation of the
proximity effect is an ac current, which appears in case of
voltage-biased junction. We show that the interplay of two above
mentioned phenomena in non-equilibrium SFS Josephson junction
gives an opportunity to implement a source of highly
spin-polarized ac current by tunnel coupling the interlayer region
to an additional electrode. The frequency of this ac current is
controlled by the voltage $V$, applied to the SFS junction. In
addition the proximity effect in the interlayer causes substantial
non-linearities in the $I-V_p$ characteristics of the ac current
flowing through the additional electrode. Here $V_p$ is the
potential of the probe electrode. The dc current flowing through
the additional electrode is also considered, although its spin
polarization is obtained to be rather weak except for some narrow
ranges of $V_p$. We also discuss the phase difference between the
ac current flowing through the probe electrode and the ac
Josephson current flowing across the SFS junction and show that it
depends on the position in the interlayer. Therefore the system
under consideration can be used as a source of phase-shifted
coherent ac currents by tunnel coupling of two probe electrodes at
different locations of the interlayer region.

Further the model and the method we use are described. We study a
voltage-biased SFS junction, where F is a diffusive weak
ferromagnet of length $d$ coupled to two identical superconducting
reservoirs. The superconductors are supposed to be diffusive and
have $s$-wave pairing. We assume the SF interfaces to be not fully
transparent and suppose that the resistance of the SF boundary
$R_g$ dominates the resistance of the ferromagnetic interlayer
$R_F$. We assume the parameter $(R_F/R_g)(\sigma_F/\sigma_s)$,
where $\sigma_F$ and $\sigma_s$ stand for conductivities of
ferromagnetic and superconducting materials respectively, to be
also small, what allows us to neglect the suppression of the
superconducting order parameter in the S leads near the interface.
In addition, a normal voltage-biased "probe" terminal is tunnel
coupled to the interlayer through a junction of a resistance $R_p
\gg R_g$.

We use the quasiclassical theory of superconductivity for
diffusive systems in terms of time-dependent Usadel equations
\cite{usadel}. The fundamental quantity for diffusive transport is
the momentum average of the quasiclassical Green's function
$\check g(x,\varepsilon, t) = \langle \check g(\bm p_f,
x,\varepsilon, t) \rangle_{\bm p_f}$. It is a $8\times8$ matrix
form in the product space of Keldysh, particle-hole and spin
variables. In general the quasiclassical Green's functions depend
on space $\bm R$, time $t$ variables and the excitation energy
$\varepsilon$. The considered problem is effectively
one-dimensional and $\bm R \equiv x$, where $x$ - is the
coordinate measured along the normal to the junction.

The quasiclassical Green's function $\check g(x,\varepsilon, t)$
satisfies the non-stationary Usadel equation, which in the
ferromagnetic region takes the form
\begin{equation}
\left[ \varepsilon \hat \tau_3 - \check \Sigma(x, \varepsilon, t),
\check g \right]_\otimes + \frac{D}{\pi}\partial_x \left( \check g
\otimes \partial_x \check g \right) = 0 \label{Usadel} \enspace ,
\end{equation}
supplemented by the normalization condition
\begin{equation}
\check g \otimes \check g = - \pi^2 \label{normal} \enspace .
\end{equation}
The product $\otimes$ of two functions of energy and time is
defined by the noncommutative convolution $A \otimes B =
e^{i(\partial_\varepsilon^A\partial_t^B-\partial_t^A\partial_\varepsilon^B)/2}A(\varepsilon,t)B(\varepsilon,t)$.
In the problem we consider the self-energy takes the form $\check
\Sigma(x, \varepsilon, t) = h \hat \sigma_3$, where $h$ is an
exchange field in the ferromagnet. $\hat \tau_i$ and $\hat
\sigma_i$ are Pauli matrices in particle-hole and spin spaces,
respectively.

In order to solve the Usadel equation it is convenient to express
quasiclassical Green's function $\check g$ in terms of Riccati
coherence functions $\hat \gamma^{R,A}$ and $\hat {\tilde
\gamma}^{R,A}$, which measure the relative amplitudes for
normal-state quasiparticle and quasihole excitations and
distribution functions $\hat x^K$ and $\hat {\tilde x}^K$. All
these functions are $2 \times 2$ matrices in spin space and depend
on $(x, \varepsilon, t)$. The corresponding expression for $\check
g$ takes the form\cite{Eschrig00}
$$
\check g^{R,A}  = \left(
\begin{array}{cc}
\hat g^{R,A} & \hat f^{R,A}  \\
\hat {\tilde f}^{R,A} & \hat {\tilde g}^{R,A}
\end{array}
\right) = \mp i \pi \check N^{R,A} \otimes
$$
\begin{equation}
\otimes \left(
\begin{array}{cc}
(1 + \hat \gamma^{R,A} \otimes \hat {\tilde \gamma}^{R,A}) & 2 \gamma^{R,A}  \\
- 2 \hat {\tilde \gamma}^{R,A} & -(1 + \hat {\tilde \gamma}^{R,A}
\hat \gamma^{R,A})
\end{array}
\right) \label{g_RA_Riccati} \enspace ,
\end{equation}
$$
\check g^K  = \left(
\begin{array}{cc}
\hat g^K & \hat f^K  \\
- \hat {\tilde f}^K & - \hat {\tilde g}^K
\end{array}
\right) = -2 i \pi \check N^R \otimes
$$
\begin{equation}
\left(\!\!\!\!
\begin{array}{cc}
(\hat x^K -\hat \gamma^R  \otimes \hat {\tilde x}^K \otimes \hat
{\tilde \gamma}^A )\!\!\!\!\! & -(\hat \gamma^R  \otimes \hat
{\tilde
x}^K -\hat x^K  \otimes \hat \gamma^A  \\
-(\hat {\tilde \gamma}^R  \otimes \hat x^K -\hat {\tilde x}^K
\otimes \hat {\tilde \gamma}^A)\!\!\!\!\! & (\hat {\tilde x}^K
-\hat {\tilde \gamma}^R \otimes \hat x^K \otimes \hat \gamma^A )
\end{array}\!\!\!
\right)\! \otimes \check N^A \label{g_K_Riccati} ,
\end{equation}
\begin{equation}
\check N^{R,A} \! = \! \left(\!\!\!\!
\begin{array}{cc}
\left( 1 - \hat \gamma^{R,A} \otimes \hat {\tilde \gamma}^{R,A} \right)^{-1}\!\!\!\!\!\! & 0 \\
0 \!\!\!\!\!\!& \left( 1 - \hat {\tilde \gamma}^{R,A} \otimes \hat
\gamma^{R,A} \right)^{-1}
\end{array}\!\!\!\!
\right) . \label{Ng}
\end{equation}

Riccati coherence and distribution functions obey Riccati-type
transport equations\cite{Eschrig00,Eschrig04}. For the considered
problem in the ferromagnetic region the equations read as follows
$$
2 \varepsilon \otimes \hat \gamma^R - [h \hat \sigma_3, \hat
\gamma^R] -
$$
\begin{equation}
-i D \left[ \partial_x^2 \hat \gamma^R + \partial_x \hat \gamma^R
\otimes \frac{\hat {\tilde f }^R}{i \pi} \otimes \partial_x \hat
\gamma^R \right] = 0 \enspace , \label{gamma_R}
\end{equation}
$$
\partial_t \hat x^K -D \left[ \partial_x \hat \gamma^R \otimes \frac{\hat {\tilde f
}^R}{i \pi} \otimes
\partial_x \hat x^K - \partial_x
\hat \gamma^R \otimes \frac{\hat {\tilde g }^K}{i \pi} \otimes
\partial_x \hat {\tilde \gamma}^A + \right.
$$
\begin{equation}
\left. +\partial_x^2 \hat x^K + \partial_x \hat x^K \otimes
\frac{\hat f^A}{i \pi} \otimes
\partial_x \hat {\tilde \gamma}^A \right] = 0 \enspace . \label{x}
\end{equation}
Particle-hole conjugation, denoted by $\tilde ~$, is defined by
the operation $\tilde a(x, \varepsilon, t) = a(x, - \varepsilon,
t)^*$. In addition to the conjugation symmetry, the coherence and
distribution functions obey the following symmetries $\hat
\gamma^A(x, \varepsilon, t) = \hat {\tilde \gamma}^R(x,
\varepsilon, t)^\dagger$, $\hat x^K(x, \varepsilon, t) = \hat
x^K(x, \varepsilon, t)^\dagger$.

Eqs.~(\ref{gamma_R}),(\ref{x}) should be solved together with the
boundary conditions at SF interfaces. As it was mentioned above we
consider the case when the dimensionless conductance of the
boundary $G = R_F/R_g \lesssim 1$, so the interface transparency
$T_{SF} \sim G(l/d) \ll 1$. Due to the smallness of the interface
transparency $T$ we can use Kupriyanov-Lukichev boundary
conditions at SF boundaries\cite{Kupriyanov}. In terms of Riccati
coherence and distribution functions they take the form
$$
\partial_x \hat \gamma^R_{l,r} = \pm \frac{G}{2 i \pi d}\left[ \hat f^R_{S;l,r} + \hat \gamma^R_{l,r}
\otimes \hat {\tilde g}^R_{S;l,r} - \right.
$$
\begin{equation}
\left. - (\hat g^R_{S;l,r} + \hat \gamma^R_{l,r} \otimes \hat
{\tilde f}^R_{S;l,r}) \otimes \hat \gamma^R_{l,r} \right] \enspace
, \label{boundary_cond_gamma}
\end{equation}
$$
\partial_x \hat x^K_{l,r} = \pm \frac{G}{2 i \pi d}\left[ \frac{1}{2}\left( \hat g^K_{S;l,r} + \hat \gamma^R_{l,r}
\otimes \hat {\tilde g}^K_{S;l,r} \otimes \hat {\tilde
\gamma}^A_{l,r} \right) - \right.
$$
\begin{equation}
\left. - \hat \gamma^R_{l,r} \otimes \hat {\tilde f}^K_{S;l,r} -
\left( \hat g^R_{S;l,r} + \hat \gamma^R_{l,r} \otimes \hat {\tilde
f}^R_{S;l,r} \right) \otimes \hat x^K_{l,r} - h.c. \right]
\enspace . \label{boundary_cond_x}
\end{equation}
Here Riccati coherence and distribution functions denoted by the
lower case symbols $l,r$ are taken at the left and right ends of
the ferromagnet. The quantities denoted by the lower case symbols
($S;l,r$) are corresponding Green's functions at the
superconducting side of the left and right SF interfaces. As it
was already mentioned above, under the condition
$(R_F/R_g)(\sigma_F/\sigma_s) \ll 1$ we can neglect the
suppression of superconducting order parameter in the leads and,
moreover, take the Green's functions at the superconducting side
of the boundaries to be equal to their bulk values, which can be
easily deduced from the expressions (\ref{g_RA_Riccati}) and
(\ref{g_K_Riccati}) using the following bulk values of Riccati
coherence and distribution functions
\begin{equation}
\hat \gamma_{S;l,r}^{R,A}= \left\{
\begin{array}{ll}
\frac{\displaystyle \Delta e^{-2 i e V_{l,r}t}}{\displaystyle
\varepsilon \pm i
\sqrt{\Delta^2-\varepsilon^2}}i \hat \sigma_2, & |\varepsilon|<\Delta \\
\frac{\displaystyle \Delta e^{-2 i e V_{l,r}t}}{\displaystyle
\epsilon + {\rm sgn \varepsilon}
\sqrt{\varepsilon^2-\Delta^2}}i \hat \sigma_2, & |\varepsilon|>\Delta \enspace , \\
\end{array}
\right. \label{gamma_asympt}
\end{equation}
\begin{equation}
\hat x_{S;l,r}^K = \left( 1-|\hat
\gamma_{l,r}^R(\varepsilon-eV_{l,r})|^2 \right)\tanh
\frac{\varepsilon -e V_{l,r}}{2 T} \label{x_asympt} \enspace ,
\end{equation}
$\Delta$ is the superconducting order parameter absolute value in
the bulk, which is assumed to be the same in the both
superconductors. $V_{l,r}$ is the electric potential in the bulk
of left (right) superconductor, so $V=V_r-V_l$ is the voltage bias
applied to the junction. $T$ is the temperature.

The electric and spin currents flowing through the probe electrode
should be found via Keldysh part of the quasiclassical Green's
function. It is convenient to calculate the currents in the probe
electrode. Given the quasiclassical Green's function $\check
g_p({\bm p}_f, \varepsilon, t)$ at the interface of the probe
electrode the corresponding expression for the electric current
reads as follows
\begin{equation}
\frac{j^{el}}{e}\! =\! \left \langle\! N_f v_{f,y}\!\!\! \int
\limits_{-\infty}^{+\infty}\!\! \! \frac{d \varepsilon}{4 \pi i}
{\rm Tr}_4 \left[ \frac{(\hat \tau_0 + \hat \tau_3)}{2} \hat
\sigma_{0} \check g^K_p({\bm p}_f, x, \varepsilon, t) \right]
\right \rangle_{{\bm p}_f}. \label{current}
\end{equation}
where $e$ is the electron charge and $\hbar = 1$ throughout the
paper. The spin current $j^{sp}/s^e$ can be calculated making use
of Eq.~(\ref{current}) with the substitution $\hat \sigma_3$ for
$\hat \sigma_0$. $s^e=1/2$ is the electron spin. In
Eq.~(\ref{current}) $v_{f,y}$ is the Fermi velocity component
normal to the junction between the interlayer and the probe
electrode, $N_f$ is the density of the states on the Fermi level
in the probe electrode, $\langle ... \rangle_{{\bm p}_f}=\int
\limits_{FS} d \Omega /4 \pi $ denotes average over the Fermi
surface.

In the first order on the transparency $T_{FP}$ of the junction
between the interlayer and the probe electrode (which is
considered to be small) the difference between the Green's
functions for an incoming ${\bm p}_f$ and outgoing ${\underline
{\bm p}}_f$ quasiparticle trajectories $\check g_p({\bm p}_f, x,
\varepsilon, t)-\check g_p({\underline {\bm p}}_f, x, \varepsilon,
t)$ entering Eq.~(\ref{current}) can be expressed in terms of the
Green's functions corresponding to the uncoupled interlayer and
probe electrode regions as follows \cite{mrs88}
$$
\check g_p({\bm p}_f, x, \varepsilon, t)-\check g_p({\underline
{\bm p}}_f, x, \varepsilon, t)=
$$
\begin{equation}
=-\frac{i T_{FP}}{2 \pi}\left[ \check g_F({\bm p}_f, x,
\varepsilon, t), \check g_p^{(0)}({\bm p}_f, \varepsilon, t)
\right]_\otimes \label{tun_boundary_cond} \enspace .
\end{equation}
Here the Green's function $\check g_F({\bm p}_f, x, \varepsilon,
t)$ in the dirty ferromagnetic interlayer is only slightly
dependent on the momentum direction on the Fermi surface and
approximately equal to its momentum average value $\check g_F(x,
\varepsilon, t)$. $[A,B]_\otimes$ means the commutator $A \otimes
B - B \otimes A $. Substituting Eq.~(\ref{tun_boundary_cond}) into
the expression for the current (\ref{current}) and taking into
accout the explicit form of the Green's function for the uncoupled
normal probe electrode $\check g_p^{(0)R,A}=\mp i \pi \hat \tau_3
\hat \sigma_0$ and $g_p^{(0)K}=- 2 i \pi \tanh
[(\varepsilon-eV_p)/2T] \hat \tau_3 \hat \sigma_0$, we get the
current flowing through this electrode:
$$
j^{el} = \frac{1}{eR_p}\int \limits_{-\infty}^\infty \frac{d
\varepsilon}{4 \pi i} {\rm Tr}_2 \hat \sigma_0 \Bigl\{\hat
g_F^K(x, \varepsilon, t) +
$$
\begin{equation}
\tanh \frac{\varepsilon-eV_p}{2T} \otimes \hat g_F^A(x,
\varepsilon, t)-\hat g_F^R(x, \varepsilon, t)\otimes \tanh
\frac{\varepsilon-eV_p}{2T} \Bigr\} \label{probe_current} \enspace
.
\end{equation}
$\hat g_F^{R,A,K}(x, \varepsilon, t)$ is the upper left part of
the interlayer Green's function $\check g_F^{R,A,K}(x,
\varepsilon, t)$ in the particle-hole space. The spin current
$j^{sp}/s^e$ can be calculated from Eq.~(\ref{probe_current}) with
the substitution $\hat \sigma_3$ for $\hat \sigma_0$. In
Eq.~(\ref{probe_current}) the resistance of the interface between
the interlayer and the probe electrode $R_p^{-1}=e^2
\int_{v_{f,y}>0} (d \Omega / 4 \pi) v_{f,y}N_{f,y} T_{FP}$.

The Green's functions $\check g_F^{R,A,K}(x, \varepsilon, t)$ are
calculated making use of the Riccati-parameterization technique
described above. The time dependence of the superconducting order
parameter in the leads $\sim \Delta e^{2 i e V t}$, which cannot
be removed by the gauge transformation in the case of
voltage-biased SFS junction, give rise to time dependence of the
Green's function in the interlayer:
\begin{equation}
\check g_F^{R,A,K}(x, \varepsilon, t)=\sum
\limits_{m=-\infty}^{\infty} \check g_m^{R,A,K}(x,
\varepsilon)e^{2 i e m V t}\label{exp_time} \enspace .
\end{equation}
If the interlayer is long in comparison with the coherence length,
all the Green's function harmonics corresponding to $m \neq 0$ are
negligible. This leads, in particular, to the suppression of the
ac Josephson effect in SFS junction. However, if the length of the
interlayer is of the order of the coherence length, the non-zero
harmonics are important and give rise not only to ac Josephson
effect, but also to ac electric and spin currents flowing through
the probe electrode. It is worth to note here that we consider
exchange fields, which are very weak as compared to the Fermi
energy $\varepsilon_F$ and calculate all the currents to zero
order of the parameter $h/\varepsilon_F$. In this approximation
the spin current across SFS junction is absent (it only appears in
the first order of this parameter). At the same time non-zero spin
current flows through the probe electrode. The qualitative reason
for this effect is the manifestation of the proximity induced
minigap, which is splitted by the exchange field, in the
essentially nonequilibrium electron distribution function in the
interlayer. This is discussed in detail below.

\begin{figure}[!tbh]
   \begin{minipage}[b]{\linewidth}
   \centerline{\includegraphics[clip=true,width=3in]{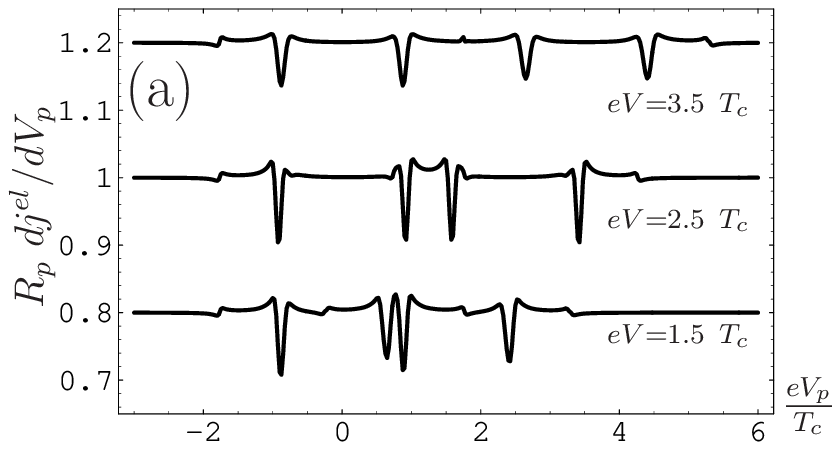}}
  \end{minipage}
 \begin{minipage}[b]{0.65\linewidth}
   \centerline{\includegraphics[clip=true,width=1.8in]{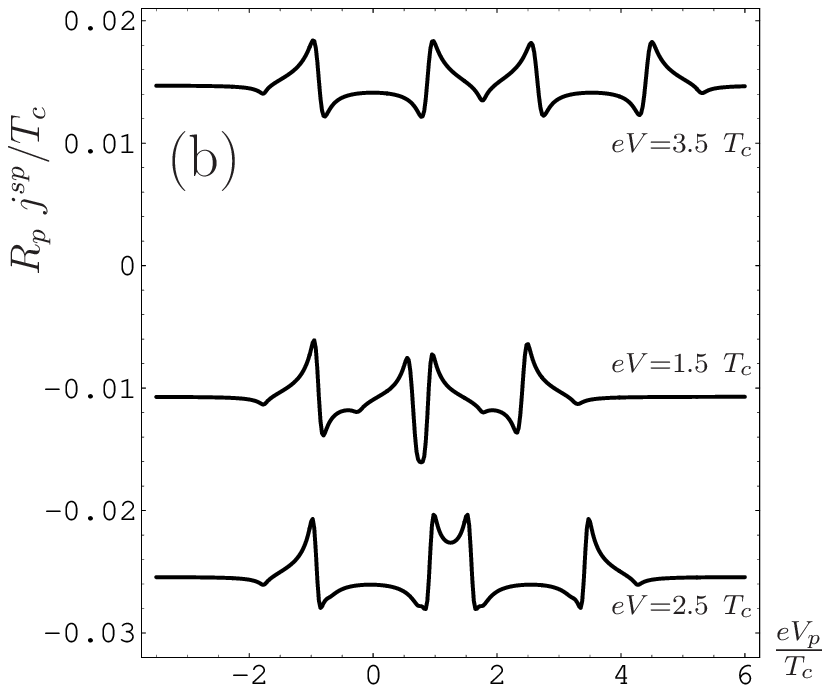}}
     \end{minipage}
  \begin{minipage}[b]{0.30\linewidth}
      \centerline{\includegraphics[clip=true,width=1.30in]{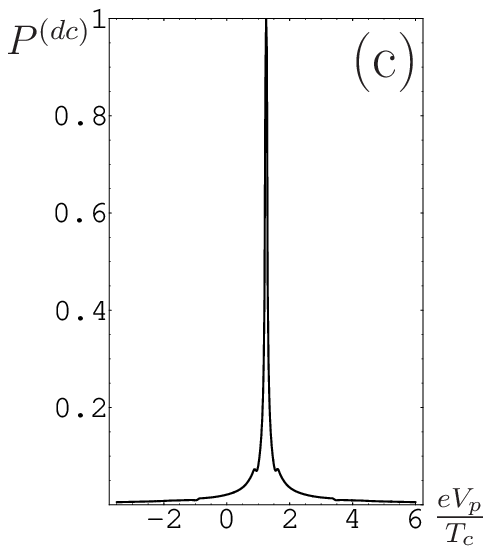}}
  \end{minipage}
     \caption{(a) The differential conductance of the dc electric current
     flowing through the probe electrode as a function of $eV_p$.
     Different curves are related to three different values of the voltage $V$,
     applied to SFS junction. The conductance is normalized to
     its asymptotic value $1/R_p$ corresponding to high
     enough $|V_p|$. The offset is for clarity. (b) The dependence of the dc spin current
     on $eV_p$ for the same three values of $V$. There is no offset. The spin current
     is measured in units of $T_c/R_p$, where $T_c$ is the critical temperature of the
     superconducting leads. (c) The degree of spin polarization $P^{(dc)}$ of the dc current for $eV=2.5 T_c$.
     For other values of the voltage $V$ the behavior of $P^{(dc)}$ is
     qualitatively the same. The other parameters of the junction are the following: $h=0.9T_c$, $d=0.94 \xi$,
     $G=0.04$. All the results presented in the paper are calculated for the temperature $T=0.01T_c$.}\label{dc_V}
\end{figure}

\begin{figure}[!tbh]
   \begin{minipage}[b]{\linewidth}
   \centerline{\includegraphics[clip=true,width=3in]{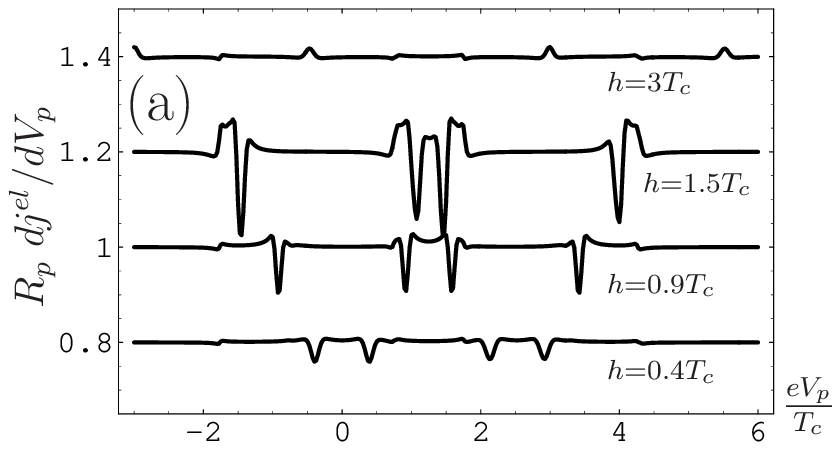}}
  \end{minipage}
 \begin{minipage}[b]{\linewidth}
   \centerline{\includegraphics[clip=true,width=3in]{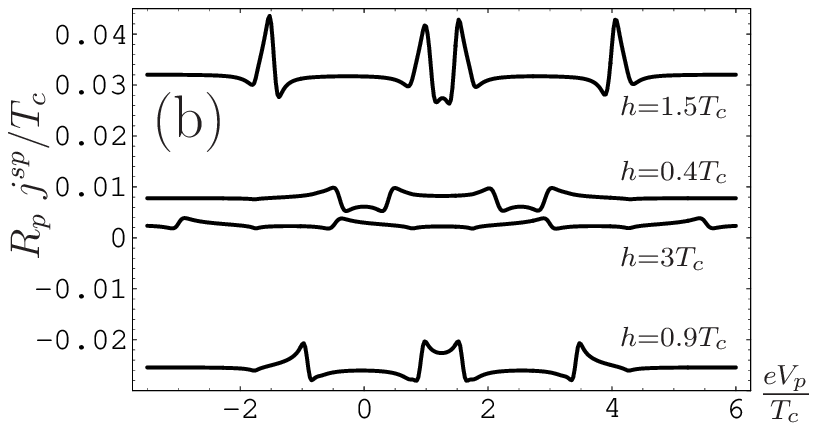}}
  \end{minipage}
   \caption{(a) The differential conductance of the dc electric current
     flowing through the probe electrode as a function of $eV_p$.
     Different curves are related to four different values of exchange field $h$.
     The offset is for clarity. (b) The dependence of the dc spin current
     on $eV_p$ for the same values of $h$. There is no offset.
     The other parameters of the junction are the following: $eV=2.5 T_c$,
     $d=0.94 \xi$ and $G=0.04$.}\label{dc_h}
\end{figure}

At first, let us focus on the behavior of the dc component of the
electric and spin currents flowing through the probe electrode.
The differential conductance $dI/dV$ of the electric current
(normalized to its asymptotic value $1/R_p$ corresponding to high
enough $|V_p|$) in dependence on $V_p$ is plotted in
Figs.~\ref{dc_V}(a) and \ref{dc_h}(a). The offset is for clarity.
Figs.~\ref{dc_V}(b) and \ref{dc_h}(b) represent the dependence of
the dc spin current on $V_p$. Fig.~\ref{dc_h} demonstrates how
these quantities are affected by the exchange field, while
Figs.~\ref{dc_V}(a)-(b) show the influence of voltage $V$ applied
to SFS junction.

The most interesting feature of the electric current is series of
dips in the differential conductance, which is seen in
Figs.~\ref{dc_V}(a) and \ref{dc_h}(a). These dips are direct
consequence of the proximity effect and reflect the spin-split
minigaps in the LDOS extended from the left interface
(corresponding to the voltages $V_p=\pm h$) and from the right one
(located at the positions $V_p=V \pm h$). The sign $\pm$ is
related to the different spin subbands. It is seen in
Fig.~\ref{dc_V}(a) that the relative positions of the dip pairs
can be adjusted by manipulating the voltage $V$ applied to SFS,
which is easily controlled experimentally. Fig.~\ref{dc_h}(a)
demonstrates that these features are the most pronounced for weak
exchange fields $h \lesssim \Delta$. For $h > \Delta$ the minigaps
are pushed out from the subgap regions of the LDOS  and convert
into obscure features. Consequently, the dips in the differential
conductance become evanescent for high enough exchange fields.

For weak exchange fields $h \lesssim \Delta$ the dc spin current,
which is represented in Figs.~\ref{dc_V}(b) and \ref{dc_h}(b), is
also a highly non-linear function of $eV_p$ if it is roughly
between $-\Delta$ and $\Delta+eV$. This is again the consequence
of the interplay between the minigaps extended from the both
interfaces of the ferromagnet. It is worth to note here that the
whole picture is symmetrical with respect to $V_p-V/2 \to
-(V_p-V/2)$. The reason is that the curves, represented in
Figs.~\ref{dc_V} and \ref{dc_h}, are obtained at $x=d/2$, that is
exactly in the middle of the interlayer. If one would calculate
the currents flowing through the probe electrode closer to one of
the interfaces, there would be an asymmetry of the corresponding
curves with respect to $V_p-V/2 \to -(V_p-V/2)$. This asymmetry
originates from the fact that the proximity features extended from
the nearest boundary dominate the proximity features, which
extended from the other one. It is seen in Fig.~\ref{dc_V}(b) that
the distance between the non-linear features in the spin current
is again can be controlled by the voltage $V$. Another non-trivial
characteristic feature is that the dc spin current tends to a
constant value at large enough $|V_p|$ instead of to be a linear
function of the voltage bias. This constant value is a
non-monotonous function of the exchange field and declines upon
increasing $h$ (just as the amplitude of the non-linear features
does). This behavior is a consequence of the manifestation of
spin-split minigap in the nonequilibrium quasiparticle
distribution function in the interlayer and is discussed
qualitatively below. However, it is obvious that full dc spin
current does not vanish upon increasing $h$ if one takes into
account the contributions of the first and the following orders of
$h/\varepsilon_F$, which are disregarded here but become essential
for larger exchange fields.

Fig.~\ref{dc_V}(c) demonstrates the degree of spin polarization of
the current as a function of $V_p$. For the dc component of the
current this quantity is defined as
\begin{equation}
P^{(dc)}=\frac{|j^{sp}|/s^e}{|j_\uparrow|+|j_\downarrow|} \equiv
\frac{|j_\uparrow-j_\downarrow|}{|j_\uparrow|+|j_\downarrow|}
\label{polarization_dc} \enspace ,
\end{equation}
where $j_{\uparrow(\downarrow)}$ is dc component of the current of
spin-up (spin-down) electrons, which can be calculated according
to Eq.~(\ref{probe_current}) with the substitution $(\hat
\sigma_0+\hat \sigma_3)/2$ ($(\hat \sigma_0-\hat \sigma_3)/2$) for
$\hat \sigma_0$. It is seen that the polarization of the dc
current is weak everywhere except for the particular narrow ranges
of $V_p$, where the electric current is small due to smallness of
the voltage bias $V_p-V/2$ applied between the interlayer and the
probe electrode.

\begin{figure}[!tbh]
   \begin{minipage}[b]{\linewidth}
   \centerline{\includegraphics[clip=true,width=3in]{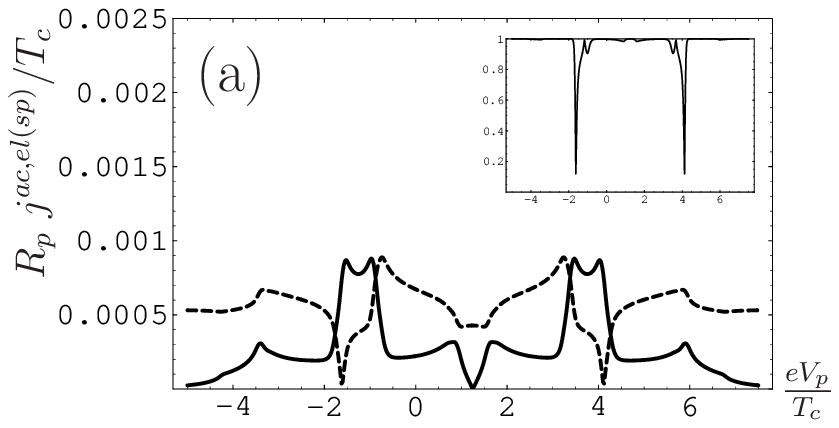}}
  \end{minipage}
 \begin{minipage}[b]{\linewidth}
   \centerline{\includegraphics[clip=true,width=3in]{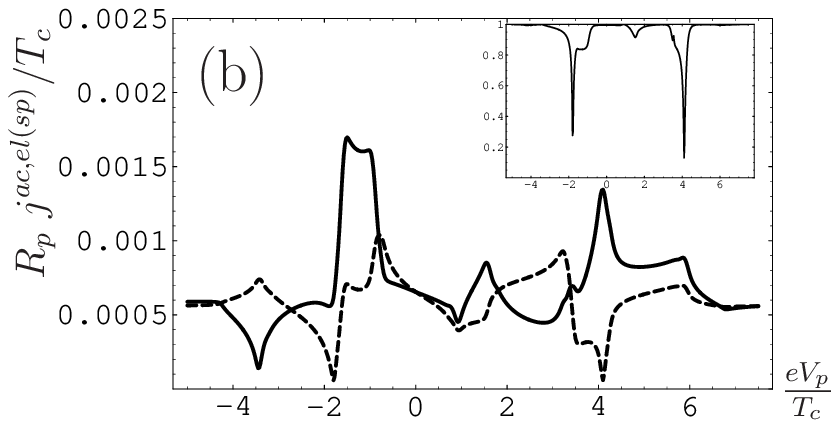}}
     \end{minipage}
       \caption{The dependencies of the ac electric (solid line)
       and ac spin (dashed line) currents on $V_p$. $h=0.9 T_c$,
       $V=2.5 T_c$, other parameters of SFS junction are the same
       as in Figs.~\ref{dc_V} and \ref{dc_h}. Panel (a) demonstrates
       the results for the case when the probe electrode is
       coupled to the interlayer at $x=d/2$, while the curves, represented in panel (b)
       are calculated at $x=0$.}
       \label{ac}
\end{figure}

Now we turn to the discussion of ac current. Fig.~\ref{ac} is a
representative example of the dependencies of ac electric and spin
currents on $V_p$. Panel (a) demonstrates the currents in the
middle of the interlayer (at $x=d/2$), while the currents at the
left interface ($x=0$) are plotted in panel (b). We only consider
the first harmonics of ac current, corresponding to $m=\pm 1$ in
Eq.~(\ref{exp_time}). The following harmonics are negligible
because we assume the transparency of SF interfaces to be low
enough, what leads to the suppression of a Green's function
component $g_{m+1}$ by a factor of $G$ with respect to $g_m$.
Therefore, the electric and spin ac currents can be expressed by
$j^{ac,el(sp)}(t)=j^{el(sp)}_1 e^{2ieVt}+j^{el(sp)}_{-1}
e^{-2ieVt}$. Then the amplitudes of these currents, which are
represented in Fig.~\ref{ac}, take the form
$|j^{ac,el(sp)}|=|j^{el(sp)}_1|+|j^{el(sp)}_{-1}|$.

Similar to the dc currents, $|j^{ac,el(sp)}|$ exhibit quite strong
non-linearities if $V_p$ is between $-(\Delta+eV)$ and
$\Delta+2eV$. The ac spin current also tends to a constant value
beyond this interval. This limiting value of the current is also a
non-monotonous function of the exchange field and declines upon
increasing $h$, as for the case of the dc spin current. However,
unlike the dc spin current, it does not contain any contributions,
which rise upon increasing $h$. The ac spin current is entirely
originated from the non-equilibrium proximity effect and,
consequently, is suppressed for the exchange fields, which are
considerably larger than $\Delta$. For $h=4.5 T_c$ the ac spin
current is suppressed by a factor $\sim 4$ with respect to the
case of $h=0.9 T_c$, illustrated in Fig.~\ref{ac}.

As concerns the limiting behavior of the ac electric current, it
is very different from the dc one: it also saturates at a constant
value instead of to be a linear function of $V_p$. This limiting
value of the ac electric current is comparable to or in most part
of cases even less than the limiting value of the spin current. In
particular, in the middle of the interlayer (at $x=d/2$) the
limiting value of the electric current is negligible (see
Fig.~\ref{ac}(a)). Therefore, the ac current is highly
spin-polarized practically for the whole parameter range we
consider. We define the polarization of the ac current by the
following way
\begin{equation}
P^{(ac)}=\frac{\langle |j^{ac,sp}(t)|P(t) \rangle}{\langle
|j^{ac,sp}(t)| \rangle} \label{polarization_ac} \enspace ,
\end{equation}
where $\langle ... \rangle$ means averaging over time and $P(t)$
is the instantaneous spin polarization of the ac current, defined
similar to the polarization of dc current
Eq.~(\ref{polarization_dc}):
\begin{equation}
P(t)=\frac{|j^{ac,sp}(t)|/s^e}{|j^{ac}_\uparrow(t)|+|j^{ac}_\downarrow(t)|}
\label{polarization_instant} \enspace .
\end{equation}

The degree of ac current spin polarization, calculated according
to Eq.~(\ref{polarization_ac}), is represented in inserts to
Figs.~\ref{ac}(a) and \ref{ac}(b) at $x=d/2$ and $x=0$,
respectively. It is worth to note here that although the ac spin
current declines upon increasing $h$, the exchange field also
suppresses the ac electric current. For this reason the spin
polarization of the ac current remains to be high even for the
case of exchange fields $h>\Delta$. Another factor, which
suppresses the proximity effect in the interlayer is considerable
increase of the voltage $V$ with respect to $\Delta$.

\begin{figure}[!tbh]
   \begin{minipage}[b]{\linewidth}
   \centerline{\includegraphics[clip=true,width=2.5in]{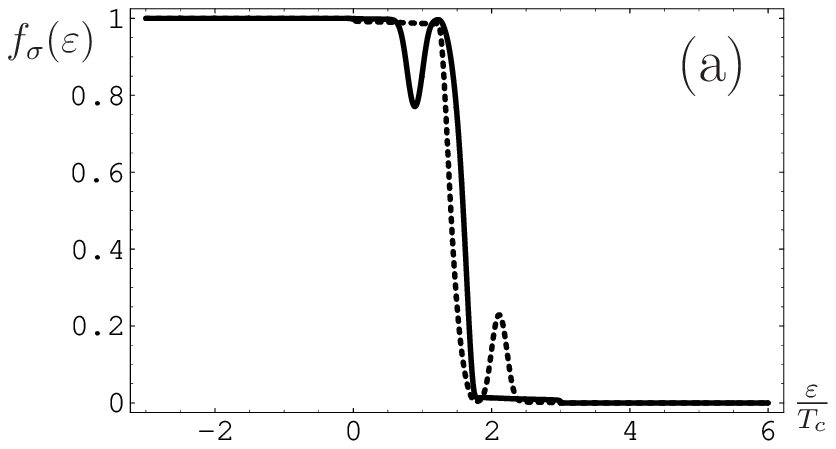}}
  \end{minipage}
 \begin{minipage}[b]{\linewidth}
   \centerline{\includegraphics[clip=true,width=2.5in]{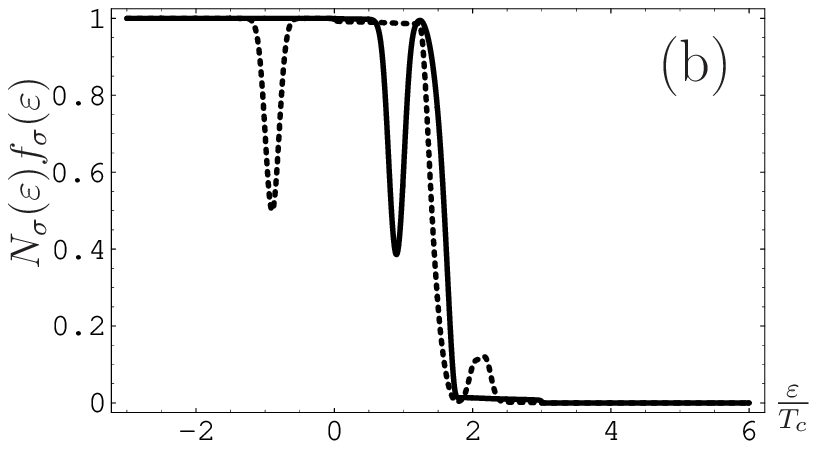}}
     \end{minipage}
       \caption{(a) The qualitative behavior of the distribution function
       $f_\sigma(\varepsilon)$ in the interlayer in dependence on $\varepsilon$.
       Solid and dashed lines are related to spin-up and spin-down
       electrons, respectively. (b) A typical behavior of the product $N_\sigma(\varepsilon)
       f_\sigma(\varepsilon)$ as a function of the quasiparticle energy.}
       \label{explanation}
\end{figure}

As it was already mentioned above, all the most essential features
of the current flowing through the probe electrode in the system
under consideration (behavior of dc spin and ac electric and spin
currents, high spin polarization of the ac current) originate from
the manifestation of spin-split proximity minigaps in the
non-equilibrium distribution function for electrons in the
interlayer. In order to get insight into qualitative behavior of
this distribution function one can make use of a balance equation.
It cannot be applicable for quantitative consideration of the
discussed problem even in case if the interlayer is short enough
to justify disregarding the inelastic relaxation processes.
Nevertheless, it can help us to clarify the qualitative mechanism
of the proximity-generated non-equilibrium effects, discussed
above. Equating flows of incoming and outgoing electrons, one
obtains the following expression for the spin-up and spin-down
distribution functions in the interlayer
\begin{equation}
f_\sigma(\varepsilon)=\frac{N^{L}_\sigma
(\varepsilon)f^L+N^{R}_\sigma (\varepsilon)f^R}{N^{L}_\sigma
(\varepsilon)+N^{R}_\sigma (\varepsilon)} \label{balance} \enspace
.
\end{equation}
Here $\sigma=\uparrow,\downarrow$ is the electron spin.
$f^L=f_F(\varepsilon)$ and $f^R=f_F(\varepsilon-eV)$ are
distribution functions in the left and right superconducting
leads. $f_F$ stands for Fermi distribution function.
$N^{L,R}_\sigma$ are local densities of states near the interface
in left and right superconductors. $N^{L,R}$ are spin-dependent
and exhibit peak features (related to the corresponding minigaps
in the ferromagnet) for subgap energy regions due to proximity
effect. Assuming density of states in the normal probe electrode
to be independent on energy and the distribution function to be
approximately step-like, the current carried by the electrons with
spin $\sigma$ through the probe electrode is proportional to the
following expression
\begin{equation}
\int \limits_{eV_p}^{\infty}d\varepsilon N_\sigma(\varepsilon,
x)f_\sigma(\varepsilon)-\int \limits_{-\infty}^{eV_p}d\varepsilon
N_\sigma(\varepsilon,
x)(1-f_\sigma(\varepsilon))\label{current_probe_approx} \enspace .
\end{equation}
The distribution function and the density of states only change
considerably within a voltage interval $[\varepsilon_1,
\varepsilon_2]$. A typical example of the distribution function
qualitative behavior in the interlayer is represented in
Fig.~\ref{explanation}(a). For $eV<\varepsilon_1$ all the
quasiparticle states  are occupied, that is $f_\sigma=1$.
Otherwise, for $eV>\varepsilon_2$ all the states are empty and
$f_\sigma=0$. The behavior of the product
$N_\sigma(\varepsilon)f_\sigma(\varepsilon)$, which is plotted in
Fig.~\ref{explanation}(b), is analogous to that one of the
distribution function. Therefore, if the potential of the probe
electrode $eV_p$ is taken to be less than $\varepsilon_1$, then
the second term in Eq.~(\ref{current_probe_approx}) can be
neglected and the current carried by spin $\sigma$ electrons is
proportional to the expression
\begin{equation}
N_{F\sigma} (\varepsilon_1-eV_p)+\int
\limits_{\varepsilon_1}^{\varepsilon_2}d\varepsilon
N_\sigma(\varepsilon, x)f_\sigma(\varepsilon)
\label{current_probe_approx1} \enspace ,
\end{equation}
where $N_{F\sigma}$ is the density of states in the ferromagnet,
corresponding to large enough quasiparticle energies. It is seen
from Eq.~(\ref{current_probe_approx1}) that  the dc electric
current is linear function of the applied voltage, as it should
be. At the same time if one neglects the difference between
$N_{F\uparrow}$ and $N_{F\downarrow}$, that is only considers the
zero order of the parameter $h/\varepsilon_F$, then the dc spin
current is only originated from the second term, which is
independent of $V_p$. This is proximity driven contribution to the
current, because if the proximity features are absent, this term
is spin and time independent, as it can be seen from
Eq.~(\ref{balance}).

On the contrary, in the presence of the proximity induced features
the product $N_\sigma(\varepsilon)f_\sigma(\varepsilon)$ is
different for spin-up and spin-down electrons in the interval
$[\varepsilon_1,\varepsilon_2]$, as it is demonstrated in
Fig.~\ref{explanation}(b), what leads to the discussed above
constant limiting behavior of the dc spin current. The limiting
constant behavior of the ac current also originates from this
term. The point is that the particular shape (especially width) of
the minigap features is fairly sensitive to the phase difference
at SFS junction. As in the nonequlibrium conditions the phase
difference is driven by the factor $2eVt$, all the features
generated by the minigap in Fig.~\ref{explanation} oscillate in
time with the corresponding frequency, giving rise to ac
contribution to the current.

\begin{figure}[!tbh]
   \begin{minipage}[b]{\linewidth}
   \centerline{\includegraphics[clip=true,width=3in]{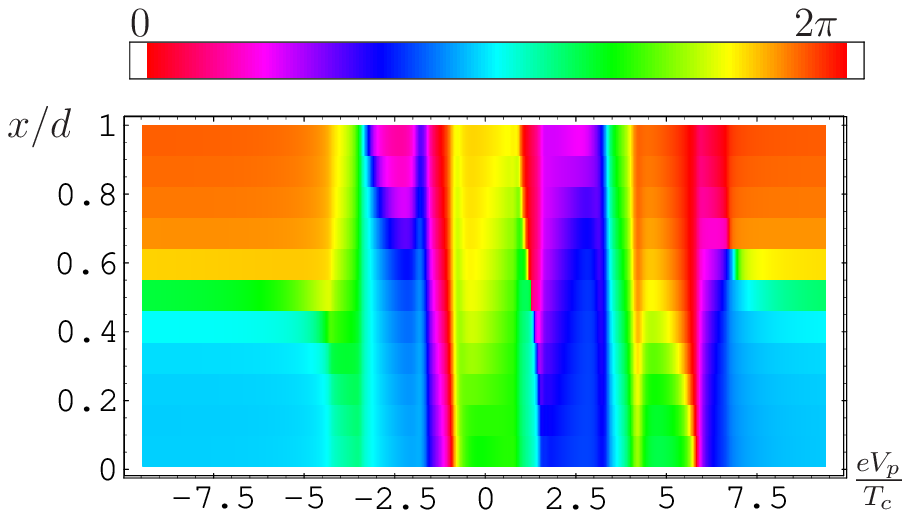}}
  \end{minipage}
        \caption{The distribution of the
ac electric current phases flowing through the probe electrode in
dependence of the coordinate $x$ across the interlayer (vertical
axis) and the potential of the probe electrode $V_p$ (horizontal
axis).}
\label{phase}
\end{figure}

Another interesting property of the ac current flowing through the
probe electrode is the possibility to obtain coherent highly
spin-polarized ac currents with a certain phase shift between
them. The point is that the values of the Green's functions $\hat
g^{R,A,K}_F$ in the interlayer, which determine the current
flowing through the probe electrode, depend on the coordinate $x$
across SFS junction. At the same time the frequency of their
oscillating part is position-independent and controlled by the
voltage $V$ applied to SFS junction. Therefore, by coupling two
probe electrode to different locations in the interlayer one can
implement a source of coherent ac currents having a certain phase
difference between them. Fig.~\ref{phase} demonstrates the
distribution of the ac electric current phases flowing through the
probe electrode in our system in dependence of the coordinate $x$
across the interlayer (vertical axis) and the potential of the
probe electrode $V_p$ (horizontal axis). The phase of the ac
current is calculated with respect to the phase of the ac electric
current, flowing through the SFS junction itself, because this
reference value is independent on $V_p$ and location in the
interlayer.

It is seen in Fig.~\ref{phase} that the phase shift close to $\pi$
can be obtained for a wide $V_p$ range by coupling the probe
electrodes symmetrically with respect to the middle of the
junction. In principle, phase shifts close to $\pi/2$ are also
reachable, although for a narrower ranges of $V_p$, as it is seen
in Fig.~\ref{phase}.

The most part of the discussed above results is pronounced in the
regime, when the value of the exchange field $h$ (measured in the
energy units) is of order of superconducting order parameter
$\Delta$. In ferromagnetic alloys like $CuNi$, which have been
intensively used by now for experimental investigation of
equilibrium properties of SFS heterostructures, the exchange field
is several times larger than $\Delta$. However, as far as we know,
the work on the creation of appropriate alloys is in progress now,
so we believe that this limit can be experimentally realized in
the nearest future. As the F layer is supposed to be an alloy, a
role of magnetic scattering may be quite
important\cite{Sellier03,Ryazanov04}. Therefore, the influence of
the magnetic scattering on the results, obtained in this paper
should be also investigated.

In conclusion, in the paper we have theoretically obtained that
voltage-biased SFS junction in the regime of the essential
proximity effect can be used as a source of highly spin-polarized
ac current by tunnel coupling the interlayer region to the
additional normal electrode. This current is driven by the
non-equilibrium proximity effect in the system and vanishes if the
proximity features are suppressed somehow. The frequency of the ac
current is controlled by the voltage $V$ applied to SFS junction.
In addition this system can implement a source of coherent ac
currents having a certain phase difference between them by
coupling two probe electrode to different locations in the
interlayer.

{\it Acknowledgments} The support by the Russian Science Support
Foundation (A.M.B.), RF Presidential Grant No.MK-4605.2007.2
(I.V.B.) and the programs of Physical Science Division of RAS are
acknowledged.


\end{document}